# Market-Based Portfolio Variance

Victor Olkhov
Independent, Moscow, Russia
victor.olkhov@gmail.com

5 July 2025.

ORCID: 0000-0003-0944-5113

## Abstract

The variance measures the portfolio risks the investors are taking. The investor, who holds his portfolio and doesn't trade his shares, at the current time can use the time series of the market trades that were made during the averaging interval with the securities of his portfolio and assess the current return, variance, and hence the current risks of his portfolio. We show how the time series of trades with the securities of the portfolio determine the time series of trades with the portfolio as a single market security. The time series of trades with the portfolio determine its return and variance in the same form as the time series of trades with securities determine their returns and variances. The description of any portfolio and any single market security is equal. The time series of the portfolio trades define the decomposition of the portfolio variance by its securities, which is a quadratic form in the variables of relative amounts invested into securities. Its coefficients themselves are quadratic forms in the variables of relative numbers of shares of its securities. If one assumes that the volumes of all consecutive deals with each security are constant, the decomposition of the portfolio variance coincides with Markowitz's (1952) variance, which ignores the effects of random trade volumes. The use of the variance that accounts for the randomness of trade volumes could help majors like BlackRock, JP Morgan, and the U.S. Fed to adjust their models, like Aladdin and Azimov, to the reality of random markets.

---

This research received no support, specific grant, or financial assistance from funding agencies in the public, commercial, or nonprofit sectors. We welcome offers of substantial support.

## 1. Introduction

The investors who hold their portfolios take the risks that are measured by the portfolio variance. The monitoring of the variance is crucial for maintaining the limitation of risks. The variance forecasting is the main tool for optimal portfolio selection. More than seventy years ago, Markowitz (1952) described the portfolio variance $\Theta(t,t_0)$ (1.2) as the quadratic form in variables $X_j(t_0)$ of relative amounts invested into the securities with the coefficients equal to the covariances $\theta_{jk}(t,t_0)$ (1.3) of the returns of the securities that compose the portfolio. This result allowed him to formulate the principles of optimal selection of the portfolio with higher returns under lower variance. Since then, portfolio selection theory has been developed a lot (Pogue, 1970; Markowitz, 1991; Rubinstein, 2002; Cochrane, 2014; Elton et al., 2014; Boyd et al., 2024). However, Markowitz's expression of the portfolio variance $\Theta(t,t_0)$ (1.2) remains unchanged.

We believe that Markowitz's (1952) result is well known and needs no additional clarifications. We follow Markowitz and consider the portfolio that was collected of $j=1,.. J$ securities in the past at time $t_0$. The mean return $R(t,t_0)$ (1.1) of the portfolio at current time $t$ takes the form:

$$R(t,t_0) = \sum_{j=1}^{J} R_j(t,t_0)\, X_j(t_0) \tag{1.1}$$

$R_j(t,t_0)$ in (1.1) denotes the mean return of security $j$ at time $t$ with respect to time $t_0$. The coefficients $X_j(t_0)$ in (1.1) denote the relative amounts invested into security $j$ at time $t_0$. All prices are adjusted to the current time $t$. Markowitz (1952) derived the portfolio variance $\Theta(t,t_0)$ (1.2) as a quadratic form in the variables of the relative amounts $X_j(t_0)$ invested into security $j$:

$$\Theta(t,t_0) = \sum_{j,k=1}^{J} \theta_{jk}(t,t_0) X_j(t_0)\, X_k(t_0) \tag{1.2}$$

The coefficients $\theta_{jk}(t,t_0)$ (1.3) in (1.2) denote the covariances of returns of securities $j$ and $k$:

$$\theta_{jk}(t,t_0) = E\left[\left(R_j(t_i,t_0) - E[R_j(t_i,t_0)]\right)\left(R_k(t_i,t_0) - E[R_k(t_i,t_0)]\right)\right] \tag{1.3}$$

The expression of the portfolio variance $\Theta(t,t_0)$ (1.2) for decades served successfully as a basis for the optimal portfolio selection and the portfolio theory as a whole.

Actually, any valuable results in economics and finance are the consequences of particular approximations and assumptions. We restudy the portfolio variance and show that the assessment of the portfolio variance $\Theta(t,t_0)$ (1.2) is correct only if one assumes that the volumes of consecutive market trades with all securities that compose the portfolio are constant during the averaging interval. Meanwhile, the time series of volumes of trades with the securities reveal their high irregularity or randomness during any reasonable interval. We derive a market-based expression of the portfolio variance $\Theta(t,t_0)$ that accounts for the randomness of the volumes of consecutive market trades with the securities. Our expression of the portfolio variance $\Theta(t,t_0)$ is also a quadratic form in the variables $X_j(t_0)$ of the relative amounts invested into securities, but its coefficients differ a lot from the covariances $\theta_{jk}(t,t_0)$ (1.3) of securities.

We highlight that our model has nothing in common with numerous studies (Karpoff, 1986; Lo and Wang, 2001; Goyenko et al., 2024) that consider the issues of trading volume in a different manner.

Market-based portfolio variance $\Theta(t,t_0)$ that accounts for the randomness of the volumes of the consecutive trades with the securities of the portfolio changes the existing methods for optimal selection and portfolio compositions. We don't study these problems here. The investors and portfolio managers can use our results to adjust their models of optimal portfolio selections.

In Section 2, we describe the portfolio that was collected by the investor of $j=1,..J$ securities in the past at time $t_0$. The investor holds his portfolio and doesn't trade his shares. To assess the average return and variance of his portfolio at current time $t$, the investor uses the time series of all market trades that were made with all securities of his portfolio during the averaging interval. We show how the investor should transform the time series of market trades with the securities of his portfolio to obtain the time series that describe the trades with his portfolio as a single market security. The time series of trades with the portfolio equally describe the return and variance of the portfolio that is composed of many securities $j=1,..J$, $J>>1$, and the portfolio that is composed of one security $J=1$.

In Section 3, we derive how the time series of values and volumes of trades with a single security $j$ of the portfolio and the time series of values and volumes of trades with the portfolio as a single security determine their average prices, returns, and the variances of prices and returns. In Section 4, we describe the decomposition of the price, return, and their variances of the portfolio by its securities. The decomposition of the portfolio variance $\Theta(t,t_0)$ by its securities is a quadratic form in the variables $X_j(t_0)$ (1.1; 1.2) of the relative amounts invested into securities. Its coefficients $\Omega_{jk}(t)$ (B.12), in turn, are quadratic forms in the variables $x_j(t_0)$ of the relative number of shares of security $j$ in the total number of shares of the portfolio. If one assumes that all volumes of trades that were made with all securities during the averaging interval are constant, the expression of the portfolio variance takes the form (1.2). The conclusion is in Section 6.

We derive some results in Appendices A and B. In App. A, we derive the market-based covariances of prices and returns of two securities. In App. B, we derive the decomposition of the variance of the portfolio by its securities. All prices are adjusted to the current time $t$.

## 2. Time series of trades with the portfolio as a single market security

We assume that at time $t_0$, the investor collected his portfolio of $j=1,..J$ market securities. The investor holds his portfolio and doesn't trade the shares of his portfolio. We denote the investor's portfolio at time $t_0$ by the number $U_j(t_0)$ and the values $C_j(t_0)$ of shares of marker securities $j=1,..J$. The prices $p_j(t_0)$ per share of each security $j$ obey trivial equations:

$$C_j(t_0) = p_j(t_0)U_j(t_0) \quad ; \quad j = 1, \dots J \tag{2.1}$$

The prices $p_j(t)$ and the values $C_j(t)$ of the shares of security $j$ can change in time $t$, but the number of shares $U_j(t_0)$ of each security $j$ in the portfolio remains constant. We denote the value $Q_\Sigma(t_0)$ and the volume $W_\Sigma(t_0)$ or the total number of shares of the portfolio at time $t_0$:

$$Q_\Sigma(t_0) = \sum_{j=1}^{J} C_j(t_0) \quad ; \quad W_\Sigma(t_0) = \sum_{j=1}^{J} U_j(t_0) \tag{2.2}$$

We define the price $s(t_0)$ (2.3) per one share of the portfolio:

$$Q_\Sigma(t_0) = s(t_0)W_\Sigma(t_0) \quad ; \quad s(t_0) = \sum_{j=1}^{J} p_j(t_0)\, x_j(t_0) \quad ; \quad x_j(t_0) = \frac{U_j(t_0)}{W_\Sigma(t_0)} \tag{2.3}$$

We determine the portfolio at time $t_0$ by its value $Q_\Sigma(t_0)$, volume $W_\Sigma(t_0)$, price $s(t_0)$, and by the set of values $C_j(t_0)$, volumes $U_j(t_0)$, and prices $p_j(t_0)$ of each security $j=1,..J$ of the portfolio. Relations (2.3) decompose the price $s(t_0)$ per share of the portfolio at time $t_0$ by the prices $p_j(t_0)$ (2.1) of its securities. The coefficients $x_j(t_0)$ define the relative numbers of the shares $U_j(t_0)$ of security $j$ in the total number of shares $W_\Sigma(t_0)$ of the portfolio.

The investor doesn't trade the shares of his portfolio but at the current time $t$ assesses the means and variances of the prices and returns of his portfolio. To do that, the investor uses the time series of market trades that were performed with all securities of his portfolio during the averaging interval. For convenience, we assume that market trades with all securities $j=1,..J$ of the portfolio occur simultaneously at the same time $t_i$ with a short time span $\varepsilon>0$ between the trades and assume that $\varepsilon$ is constant and the same for the trades with all securities $j=1,..J$. For each averaging time interval $\Delta$, the number of market trades during the interval $\Delta$ is finite $i=1,..N$. For the current time $t$ we denote the time averaging interval $\Delta$ (2.4) and consider the $N$ terms of the time series of market trades at time $t_i$ during $\Delta$:

$$\Delta = \left[t - \frac{\Delta}{2};\, t + \frac{\Delta}{2}\right] \quad ; \quad t_{i+1} = t_i + \varepsilon \in \Delta \quad ; \quad i = 1, \dots N \quad ; \quad N \cdot \varepsilon = \Delta \quad ; \quad \varepsilon > 0 \tag{2.4}$$

We assume that during $\Delta$ (2.4), $N$ trades were performed with each security $j=1,..J$ of the portfolio. Each trade with the value $C_j(t_i)$ and volume $U_j(t_i)$ at time $t_i$ during $\Delta$ (2.4) defines the price $p_j(t_i)$ (2.5) of security $j$:

$$C_j(t_i) = p_j(t_i)U_j(t_i) \quad ; \quad t_i \in \Delta \quad ; \quad i = 1, \dots N \; ; \quad j = 1, ..J \tag{2.5}$$

The investor can assess the total values $C_{\Sigma j}(t)$ and volumes $U_{\Sigma j}(t)$ of trades that were made with securities $j=1,..J$ during $\Delta$ as:

$$C_{\Sigma j}(t) = \sum_{i=1}^{N} C_j(t_i) = N \cdot C_j(t;1) \quad ; \quad U_{\Sigma j}(t) = \sum_{i=1}^{N} U_j(t_i) = N \cdot C_j(t;1) \quad ; \quad j = 1,..J \tag{2.6}$$

Functions $C_j(t;1)$ and $U_j(t;1)$ in (2.6) denote the average values and volumes of trades with security $j$ of the portfolio during $\Delta$ (2.4). We highlight that the changes of the scale $\lambda$ of the values $C_j(t_i)$ and volumes $U_j(t_i)$ of trades with security $j$ during $\Delta$ (2.4) don't change the statistical properties of the price $p_j(t_i)$. Let us for each security $j=1,..J$ of the portfolio, define the scale $\lambda_j$

$$\lambda_j = \frac{U_j(t_0)}{U_{\Sigma j}(t)} \tag{2.7}$$

For $\lambda_j$ (2.7) we define the normalized values $c_j(t_i)$ and volumes $u_j(t_i)$ of trades (2.8)

$$c_j(t_i) = \lambda_j \cdot C_j(t_i) \quad ; \quad u_j(t_i) = \lambda_j \cdot U_j(t_i) \tag{2.8}$$

The changes of scale (2.8) transform the equations (2.5) into (2.9):

$$c_j(t_i) = p_j(t_i)\, u_j(t_i) \quad or \quad \lambda_j \cdot C_j(t_i) = p_j(t_i)\, \lambda_j \cdot U_j(t_i) \tag{2.9}$$

It is obvious that random values $C_j(t_i)$, $c_j(t_i)$, and volumes $U_j(t_i)$, $u_j(t_i)$ (2.8; 2.9) define the same statistical properties of price $p_j(t_i)$ of security $j$. The change of scales (2.8) guarantees that the total normalized volume $u_{\Sigma j}(t)$ (2.10) of trades with each security $j=1,..J$ of the portfolio during $\Delta$ (2.4) equals the number of shares $U_j(t_0)$ of security $j$ of the portfolio:

$$u_{\Sigma j}(t) = \sum_{i=1}^{N} u_j(t_i) = \frac{U_j(t_0)}{U_{\Sigma j}(t)} \sum_{i=1}^{N} U(t_i) = U_j(t_0) \tag{2.10}$$

For each security $j$, the time series of normalized values $c_j(t_i)$ and volumes $u_j(t_i)$ (2.8) describe the trades with precisely $U_j(t_0)$ shares of the portfolio during $\Delta$ (2.4). Let us at time $t_i$ sum the trades with all securities $j=1,..J$ of the portfolio and determine the volumes $W(t_i)$ and values $Q(t_i)$ (2.13) of the trades with the portfolio a single security:

$$Q(t_i) = \sum_{j=1}^{J} c_j(t_i) \quad ; \quad W(t_i) = \sum_{j=1}^{J} u_j(t_i) \tag{2.11}$$

The relations (2.11) replace the initial time series of the values $C_j(t_i)$ and volumes $U_j(t_i)$ (2.5) of trades with securities $j=1,..J$ with the time series (2.11) that describe the values $Q(t_i)$ and volumes $W(t_i)$ of trades with the portfolio as a single market security. The equation (2.12) determines the portfolio price $s(t_i)$ at time $t_i$ during $\Delta$:

$$Q(t_i) = s(t_i)\, W(t_i) \quad ; \quad t_i \in \Delta \quad ; \quad i = 1, \dots N \tag{2.12}$$

The total volume of trades $W_\Sigma(t)$ (2.13) during $\Delta$ is equal to the number of shares $W_\Sigma(t_0)$ (2.2) of the portfolio at time $t_0$:

$$W_\Sigma(t) = \sum_{i=1}^{N} W(t_i) = \sum_{j=1}^{J} \sum_{i=1}^{N} u_j(t_i) = \sum_{j=1}^{J} U_j(t_0) = W_\Sigma(t_0) \quad ; \quad W(t;1) = \frac{1}{N} W_\Sigma(t_0) \tag{2.13}$$

$W(t;1)$ in (2.13) denotes the average volume of trades of the portfolio as a single market security during $\Delta$ (2.4). We highlight, that $W(t;1)$ is constant and is determined by the total number of shares $W_\Sigma(t_0)$ (2.2) of the portfolio at time $t_0$. The total value $Q_\Sigma(t)$ (2.14) of trades with the portfolio during $\Delta$ is determined by current prices of the securities:

$$Q_\Sigma(t)=\sum_{i=1}^{N}Q(t_i)=\sum_{j=1}^{J}\sum_{i=1}^{N}c_j(t_i)=\sum_{j=1}^{J}\sum_{i=1}^{N}p_j(t_i)u_j(t_i)\quad;\quad Q(t;1)=\frac{1}{N}Q_\Sigma(t) \qquad (2.14)$$

$Q(t;1)$ in (2.14) denotes the average value of trades of the portfolio as a single market security during $\Delta$ (2.4). The total values $C_{\Sigma j}(t)$ and volumes $U_{\Sigma j}(t)$ or their averages $C_j(t;1)$ and $U_j(t;1)$ (2.6) of trades with the securities of the portfolio during $\Delta$ (2.4) and the total value $Q_\Sigma(t)$ (2.14) and volume $W_\Sigma(t)$ (2.13) of trades with the portfolio during $\Delta$ (2.4) allow us alike (2.3) to define the average price $p_j(t)$ of security $j$ and the average price $s(t)$ of the portfolio (2.15):

$$C_{\Sigma j}(t)=p_j(t)U_{\Sigma j}(t)\quad;\quad C_j(t;1)=p_j(t)U_j(t;1)\quad;\quad Q_\Sigma(t)=s(t)\,W_\Sigma(t) \qquad (2.15)$$

The time series (2.11; 2.12) describe the values $Q(t_i)$, volumes $W(t_i)$, and prices $s(t_i)$ of the trades with shares of the portfolio in the same way as the time series of the values $C_j(t_i)$, volumes $U_j(t_i)$, and prices $p_j(t_i)$ (2.5) describe trades of each of the market securities $j=1,..J$. The total normalized volume $u_{\Sigma j}(t)$ (2.10) of trades with each security $j$ equals the number of shares $U_j(t_0)$ of that security in the portfolio at time $t_0$. Thus, the relations (2.10; 2.11; 2.13) prove that the time series of the volumes $W(t_i)$ (2.11) of trades of the portfolio as a single security during $\Delta$ (2.4) precisely conform to the number of shares $U_j(t_0)$ of each security $j$.

## 3. Market-based return and variance of the portfolio as a single security

The time series of the values $Q(t_i)$ and volumes $W(t_i)$ (2.11-2.15) of trades with the portfolio as a single market security determine its return and variance, as the time series of the values $C_j(t_i)$ and volumes $U_j(t_i)$ (2.5; 2.6) of trades with security $j$ determine the return and variance of market security $j$. The total value $c_{\Sigma j}(t)$ (3.1) of trades with security $j$ during $\Delta$ (2.4) equals:

$$c_{\Sigma j}(t)=\sum_{i=1}^{N}c_j(t_i)=\frac{U_j(t_0)}{U_{\Sigma j}(t)}\sum_{i=1}^{N}C_j(t_i)=\frac{U_j(t_0)}{U_{\Sigma j}(t)}C_{\Sigma j}(t) \qquad (3.1)$$

We define the average price $p_j(t)$ (3.2; 3.3) of security $j$ at the current time $t$ during $\Delta$ (2.4):

$$c_{\Sigma j}(t)=p_j(t)\,u_{\Sigma j}(t)\quad;\quad p_j(t)=\frac{c_{\Sigma j}(t)}{u_{\Sigma j}(t)}=\frac{C_{\Sigma j}(t)}{U_{\Sigma j}(t)}=\frac{c_j(t;1)}{u_j(t;1)} \qquad (3.2)$$

$$p_j(t)=\frac{1}{u_{\Sigma j}(t)}\sum_{i=1}^{N}p_j(t_i)\,u_j(t_i)=\frac{1}{U_{\Sigma j}(t)}\sum_{i=1}^{N}p_j(t_i)\,U_j(t_i) \qquad (3.3)$$

The average price $p_j(t)$ (3.2; 3.3) of $U_j(t_0)$ shares of security $j$ during $\Delta$ (2.4) takes the form of volume weighted average price (VWAP) (Berkowitz et al., 1988; Duffie and Dworczak, 2021). In (3.2) we denote the average normalized value $c_j(t;1)$ and volume $u_j(t;1)$ (3.4) of trades of security $j$ during $\Delta$ (2.4).

$$c_j(t;1) = \frac{1}{N}\sum_{i=1}^{N} c_j(t_i) \quad ; \quad u_j(t;1) = \frac{1}{N}\sum_{i=1}^{N} u_j(t_i) \tag{3.4}$$

At time $t_i$, we define the instant return $R_j(t_i,t_0)$ (3.5) of security $j$ with respect to time $t_0$:

$$R_j(t_i,t_0) = \frac{p_j(t_i)}{p_j(t_0)} \tag{3.5}$$

We use the so-called gross return $R_j(t_i,t_0)$ (3.5) instead of the usual definition of return $r_j(t_i,t_0)$:

$$r_j(t_i,t_0) = R_j(t_i,t_0) - 1 = \frac{p_j(t_i)}{p_j(t_0)} - 1 \tag{3.6}$$

The variances of both definitions of return are the same. From (3.2; 3.3), obtain the average return $R_j(t,t_0)$ of security $j$ at time $t$ during $\Delta$ (2.4):

$$R_j(t,t_0) = \frac{p_j(t)}{p_j(t_0)} = \frac{p_j(t) u_{\Sigma j}(t)}{p_j(t_0) u_{\Sigma j}(t)} = \frac{c_{\Sigma j}(t)}{p_j(t_0) U_j(t_0)} = \frac{c_{\Sigma j}(t)}{C_j(t_0)} = \frac{U_j(t_0)}{U_{\Sigma j}(t)} \frac{C_{\Sigma j}(t)}{C_j(t_0)} \tag{3.7}$$

One can present the average return $R_j(t,t_0)$ (3.8) as weighted by the normalized volumes $u_j(t_i)$ (2.8) or by the volumes $U_j(t_i)$ (2.5; 2.6) of trades with security $j$:

$$R_j(t,t_0) = \frac{1}{u_{\Sigma j}(t)} \sum_{i=1}^{N} R_j(t_i,t_0)\, u_j(t_i) = \frac{1}{U_{\Sigma j}(t)} \sum_{i=1}^{N} R_j(t_i,t_0)\, U_j(t_i) \tag{3.8}$$

We derive the market-based variance $\phi_j(t)$ (3.9) of price $p_j(t_i)$ of security $j$ that accounts for the randomness of the volumes of trades during $\Delta$ (2.4) in (A.11):

$$\phi_j(t) = E_m\left[\left(p_j(t_i) - p_j(t)\right)^2\right] = \frac{\psi_j^{\,2}(t) - 2\,\varphi_j(t) + \chi_j^2(t)}{1+\chi_j^2(t)}\, p_j^2(t) \tag{3.9}$$

The market-based variance $\theta_j(t,t_0)$ (3.10) of returns $R_j(t_i,t_0)$ (3.5) of security $j$ follows from (3.9)

$$\theta_j(t,t_0) = E_m\left[\left(R_j(t_i,t_0) - R_j(t,t_0)\right)^2\right] = \frac{\phi_j(t)}{p_j^2(t_0)} = \frac{\psi_j^{\,2}(t) - 2\,\varphi_j(t) + \chi_j^2(t)}{1+\chi_j^2(t)}\, R_j^2(t,t_0) \tag{3.10}$$

The functions $\psi_j(t)$ and $\chi_j(t)$ (see A.7; A.8; B.2;B.4) denote the coefficients of variation of the random values $C_j(t_i)$ and volumes $U_j(t_i)$ of trades with security $j$ during $\Delta$ (2.4). Coefficient of variation $\chi_j(t)$ of trade volume describes its range of fluctuations, the range of the randomness. If $\chi_j(t)=0$, all trade volumes during $\Delta$ are constant. If $\chi_j(t)>0$, one should account for the influence of random trade volumes. We give the definitions of functions $\psi_j(t)$, $\varphi_j(t)$, and $\chi_j(t)$ in (A7; A.8). The relations (3.9; 3.10) describe the market-based variances of prices and returns of security $j$ of the portfolio that account for the randomness of the volumes $U_j(t_i)$ of trades with security $j$ during $\Delta$. If one assumes that the volumes $U_j(t_i)$ of trades with security $j$ during $\Delta$ (2.4) are constant and hence $\chi_j(t)=0$, then the variance of price $\phi_j(t)$ (3.9) and the variance $\theta_j(t,t_0)$ (3.10) of security $j$ take the usual form.

If $U_j(t_i)=const$, then the VWAP $p_j(t)$ (3.2; 3.3) takes the conventional form:

$$p_j(t) = \frac{1}{U_{\Sigma j}(t)} \sum_{i=1}^{N} p_j(t_i)\, U_j(t_i) \;\rightarrow\; p_j(t) = \frac{1}{N} \sum_{i=1}^{N} p_j(t_i) \tag{3.11}$$

The average return $R_j(t,t_0)$ (3.8) of security $j$ during $\Delta$ (2.4) takes the form (3.12):

$$R_j(t,t_0) = \frac{1}{U_{\Sigma j}(t)} \sum_{i=1}^{N} R_j(t_i,t_0)\, U_j(t_i) \;\rightarrow\; R_j(t,t_0) = \frac{1}{N} \sum_{i=1}^{N} R_j(t_i,t_0) \qquad (3.12)$$

The variance $\phi_j(t)$ (3.9) of prices $p_j(t_i)$ of security $j$ takes the form (3.13)

$$\phi_j(t) = \frac{1}{N} \sum_{i=1}^{N} \left(p_j(t_i) - p_j(t)\right)^2 \qquad (3.13)$$

The variance $\theta_j(t,t_0)$ (3.10) of return $R_j(t_i,t_0)$ of security $j$ takes the form (3.14):

$$\theta_j(t,t_0) = \frac{1}{N} \sum_{i=1}^{N} \left(R_j(t_i,t_0) - R_j(t,t_0)\right)^2 \qquad (3.14)$$

We underline that the expressions of the variance $\phi_j(t)$ (3.13) of prices and the variance $\theta_j(t,t_0)$ (3.14) of return $R_j(t_i,t_0)$ (3.5) of security $j$ result from the assumption that all volumes of trades with security $j$ during $\Delta$ (2.4) are constant $U_j(t_i)=const$.

The above description shows how the average price, return, and their variances of security $j$ are expressed by the time series of normalized values $c_j(t_i)$ and volumes $u_j(t_i)$ (2.8) or equally by the time series of values $C_j(t_i)$ and volumes $U_j(t_i)$ (2.5) of trades with security $j$ during $\Delta$. That presents the evidence that the value $Q_\Sigma(t_0)$, volume $W_\Sigma(t_0)$, and price $s(t_0)$ (2.2; 2.3) that describe the initial state of the portfolio at time $t_0$ and the time series of the values $Q(t_i)$, volumes $W(t_i)$, and prices $s(t_i)$ (2.11-2.15) of trades at time $t_i$ with the portfolio as a single market security determine the return and variance of the portfolio completely in the same form as (3.5-3.10). That is the result of the transformation of the time series of the values $C_j(t_i)$ and volumes $U_j(t_i)$ of trades with the securities $j=1,..J$ of the portfolio into the time series that describe the values $Q(t_i)$ and volumes $W(t_i)$ of trades with the portfolio as a single security.

The simple substitutions (3.15) of variables:

$$C_j(t_i) \rightarrow Q(t_i) \;\; ; \quad U_j(t_i) \rightarrow W(t_i) \;\; ; \quad p_j(t_i) \rightarrow s(t_i) \; ; \; p_j(t) \rightarrow s(t) \qquad (3.15)$$

give the portfolio price, return, and their variances. From (2.15), obtain:

$$s(t) = \frac{Q_\Sigma(t)}{W_\Sigma(t)} = \frac{Q_\Sigma(t)}{W_\Sigma(t_0)} = \frac{1}{W_\Sigma(t_0)} \sum_{i=1}^{N} s(t_i)\, W(t_i) \qquad (3.16)$$

$$R(t,t_0) = \frac{s(t)}{s(t_0)} = \frac{Q_\Sigma(t)}{Q_\Sigma(t_0)} = \frac{1}{W_\Sigma(t_0)} \sum_{i=1}^{N} R(t_i,t_0) W(t_i) \quad ; \quad R(t_i,t_0) = \frac{s(t_i)}{s(t_0)} \qquad (3.17)$$

$$\Phi(t) = \frac{\psi^2(t) - 2\,\varphi(t) + \chi^2(t)}{1+\chi^2(t)}\, s^2(t) \qquad (3.18)$$

$$\Theta(t,t_0) = \frac{\Phi(t)}{s^2(t_0)} = \frac{\psi^2(t) - 2\,\varphi(t) + \chi^2(t)}{1+\chi^2(t)}\, R^2(t,t_0) \qquad (3.19)$$

Functions $\psi(t)$ and $\chi(t)$ denote coefficients of variation of random values $Q(t_i)$ and volumes $W(t_i)$ of trades with the portfolio (B.2; B.3), and the function $\varphi(t)$ (B.4) denotes the ratio of the covariance of values and volumes of the portfolio to their mean values $Q(t;1)$ and $W(t;1)$.

The above expressions describe the market-based mean price $s(t)$ (3.16), the average return $R(t,t_0)$ (3.17), the price variance $\Phi(t)$ (3.18), and the variance $\Theta(t,t_0)$ (3.19) of returns of the portfolio during $\Delta$. These market-based expressions account for the randomness of the volumes $W(t_i)$ (2.11) of trades with the portfolio as a single security. The randomness of the volumes $W(t_i)$ is determined by random properties of the volumes $U_j(t_i)$ of trades with securities $j=1,..J$ of the portfolio. It is obvious that if all volumes $U_j(t_i)$ of trades with all securities that compose the portfolio are assumed constant during $\Delta$, the expressions (3.16-3.19) take simple forms:

$$s(t) = \frac{1}{W_\Sigma(t)} \sum_{i=1}^{N} s(t_i)\, W(t_i) \;\rightarrow\; s(t) = \frac{1}{N} \sum_{i=1}^{N} s(t_i) \tag{3.20}$$

$$R(t,t_0) = \frac{1}{N} \sum_{i=1}^{N} R(t_i,t_0) \tag{3.21}$$

$$\Phi(t) = \frac{1}{N} \sum_{i=1}^{N} \left(s(t_i) - s(t)\right)^2 \tag{3.22}$$

$$\Theta(t,t_0) = \frac{1}{N} \sum_{i=1}^{N} \left(R(t_i,t_0) - R(t,t_0)\right)^2 \tag{3.23}$$

The market-based expressions (3.16-3.19) of average price $s(t)$, return $R(t,t_0)$, price variance $\Phi(t)$, and return variance $\Theta(t,t_0)$ of the portfolio coincide with the similar market-based expressions for a single security. However, the investors search for the optimal compositions of the portfolio by the set of $j=1,..J$ securities and need the dependence of the portfolio properties on the properties of its securities. Actually, the time series of the values $Q(t_i)$, volumes $W(t_i)$, and prices $s(t_i)$ (2.11-2.15) allow the investor to present the decomposition of the average price $s(t)$ (3.16), return $R(t,t_0)$ (3.17), price variance $\Phi(t)$ (3.18), and return variance $\Theta(t,t_0)$ (3.19) of the portfolio by its securities.

## 4. Decomposition of the portfolio variance by its securities

We remind readers that the investor holds his portfolio and doesn't trade his shares. The investor transforms the time series of the values $C_j(t_i)$ and volumes $U_j(t_i)$ of trades with securities $j=1,..J$ of his portfolio that were made during $\Delta$ (2.4) and derives the time series of the values $Q(t_i)$, volumes $W(t_i)$, and prices $s(t_i)$ (2.11-2.15) of trades with the portfolio as a single market security.

*4.1 Decomposition of the mean price s(t):*

The decomposition of the mean price $s(t)$ (3.16) of the portfolio by the mean prices $p_j(t)$ (3.2; 3.3) of the securities takes the form:

$$s(t) = \frac{Q_\Sigma(t)}{W_\Sigma(t_0)} = \frac{1}{W_\Sigma(t_0)} \sum_{i=1}^{N} Q(t_i) = \sum_{j=1}^{J} \left[\frac{1}{U_j(t_0)} \sum_{i=1}^{N} p_j(t_i) U_j(t_i)\right] \frac{U_j(t_0)}{W_\Sigma(t_0)} \tag{4.1}$$

The decomposition of the mean price of the portfolio by the mean prices of its securities:

$$s(t) = \sum_{j=1}^{J} p_j(t)\; x_j(t_0) \quad ; \quad x_j(t_0) = \frac{U_j(t_0)}{W_\Sigma(t_0)} \tag{4.2}$$

The coefficients $x_j(t_0)$ in (4.2) describe the relative numbers of shares $U_j(t_0)$ of the security $j$ in the total number $W_\Sigma(t_0)$ of shares of the portfolio.

*4.2 Decomposition of the return $R(t,t_0)$:*

We use (3.17) and (4.2) and obtain the decomposition of portfolio return $R(t,t_0)$ by securities:

$$R(t,t_0)=\frac{s(t)}{s(t_0)}=\sum_{j=1}^{J}\frac{p_j(t)}{s(t_0)}\,x_j(t_0)=\sum_{j=1}^{J}\frac{p_j(t)}{p_j(t_0)}\,\frac{p_j(t_0)}{s(t_0)}x_j(t_0)\ =\sum_{j=1}^{J}R_j(t,t_0)\ X_j(t_0)$$

Hence, obtain:

$$R(t,t_0)=\sum_{j=1}^{J}R_j(t,t_0)\,X_j(t_0)\quad;\quad X_j(t_0)=\frac{p_j(t_0)}{s(t_0)}x_j(t_0)=\frac{p_j(t_0)U_j(t_0)}{s(t_0)W_\Sigma(t_0)}\tag{4.3}$$

The functions $X_j(t_0)$ in (4.3) define the relative amount invested into security $j$ of the portfolio at time $t_0$. The decomposition of return $R(t,t_0)$ (4.3) coincides with (1.1):

$$X_j(t_0)=\frac{p_j(t_0)U_j(t_0)}{s(t_0)W_\Sigma(t_0)}=\frac{C_j(t_0)}{Q_\Sigma(t_0)}\tag{4.3}$$

*4.3 Decomposition of the variance $\Phi(t)$ of prices:*

In Appendix B, we derive the decomposition of the variance $\Phi(t)$ (3.18) of prices $s(t_i)$ of the portfolio by its securities $j=1,..J$ (B11; B.19) that takes the form (4.4; 4.5)

$$\Phi(t)=\sum_{j,k=1}^{J}\Omega_{jk}(t)\,p_j(t)p_k(t)\,x_j(t_0)x_k(t_0)\tag{4.4}$$

$$\Omega_{jk}(t)=\frac{\psi_{jk}(t)-2\omega_j(t)+\chi^2(t)}{1+\chi^2(t)}=1+\frac{1}{1+\chi^2(t)}\left[\psi_{jk}(t)-2\omega_j(t)-1\right]\tag{4.5}$$

The decomposition of the variance $\Phi(t)$ (4.4) of prices $s(t_i)$ of the portfolio by its securities is a quadratic form in the variables of the relative numbers $x_j(t_0)$ (4.2) of shares $U_j(t_0)$ of the security $j$ in the total number $W_\Sigma(t_0)$ of shares of the portfolio. Its coefficients $\Omega_{jk}(t)$ depend on $\chi^2(t)$ that has meaning of the square of the coefficient of variance (B.2) of the random volumes $W(t_i)$ of trades with the portfolio as a single security during $\Delta$ (2.4). In Appendix B, we derive that $\chi^2(t)$ (B.10; 4.6) is a quadratic form in variables $x_j(t_0)$ (4.2) with coefficients $\chi_{jk}(t)$ (A.8):

$$\chi^2(t)=\sum_{j,k=1}^{J}\chi_{jk}(t)\ x_j(t_0)\cdot x_k(t_0)\tag{4.6}$$

The functions $\psi_{jk}(t)$ (A.8) are the covariances between the values $C_j(t_i)$ of trade with security $j$ and the values $C_j(t_i)$ of security $k$ that are normalized to their average values $C_j(t;1)$ and $C_k(t;1)$ (2.6) during $\Delta$ (2.4). The functions $\omega_j(t)$ (B.17) have the meaning of the covariances between the values $C_j(t_i)$ of trades with security $j$ and the volumes $W(t_i)$ (2.11) of trades with the portfolio as a single market security normalized to their average values $C_j(t;1)$ (2.6) and $W(t;1)$ (2.13). If one assumes that all volumes $U_j(t_i)$ of trades with all securities $j=1,..J$ of the portfolio during $\Delta$ (2.4) are constant, and hence the volumes $W(t_i)$ of trades with the portfolio are also constant, then $\chi(t)=0$. For this assumption:

$$\Omega_{jk}(t)p_j(t)p_k(t)=\psi_{jk}(t)p_j(t)p_k(t)=\sigma_{jk}(t)\tag{4.7}$$

The $\sigma_{jk}(t)$ in (4.7) denotes usual covariances (4.8) of prices $p_j(t_i)$ and $p_k(t_i)$ of securities $j$ and $k$:

$$\sigma_{jk}(t) = \frac{1}{N} \sum_{i=1}^{N} \left(p_j(t_i) - p_j(t_i)\right)\left(p_k(t_i) - p_k(t_i)\right) \tag{4.8}$$

From (3.18) and (4.7; 4.8), obtain the decomposition of the variance $\Phi(t)$ (4.9) of prices of the portfolio by its securities:

$$\Phi(t) = \psi^2(t)s^2(t) = \sum_{j,k=1}^{J} \sigma_{jk}(t)\, x_j(t_0)x_k(t_0) \tag{4.9}$$

*4.4 Decomposition of the variance $\Theta(t,t_0)$ of returns:*

We derive the decomposition of the variance $\Theta(t,t_0)$ (3.19) of returns of the portfolio by its securities in Appendix B. It takes the form (B.21):

$$\Theta(t,t_0) = \sum_{j,k=1}^{J} \Omega_{jk}(t)\, R_j(t,t_0)R_k(t,t_0)X_j(t_0)X_k(t_0) \tag{4.10}$$

The variance $\Theta(t,t_0)$ (4.10) looks alike to Markowitz's expression (1.2), but the coefficients $\Omega_{jk}(t,t_0)R_j(t,t_0)R_k(t,t_0)$ (3.12; 4.5) differs a lot from the covariances $\theta_{jk}(t,t_0)$ (A.10) of returns of securities $j$ and $k$.

The decomposition of the variance $\Theta(t,t_0)$ (4.10) of returns of the portfolio is a quadratic form in the variables $X_j(t_0)$ (4.2) of the relative amounts invested into security $j$. The coefficients $\Omega_{jk}(t,t_0)R_j(t,t_0)R_k(t,t_0)$ (4.5), in their turn, depend on $\chi^2(t)$ (4.6), which is a quadratic form in variables $x_j(t_0)$ (4.2) with coefficients $\chi_{jk}(t)$ (A.8).

The expression of market-based variance $\Theta(t,t_0)$ (4.10) differs from Markowitz's variance $\Theta(t,t_0)$ (1.2). The only cause of these distinctions is the impact of random volumes of market trades $U_j(t_i)$ with the securities $j=1,...J$ of the portfolio during $\Delta$ (2.4).

In Appendix B, we show that if one assumes that all volumes $U_j(t_i)$ of trades with all securities $j=1,...J$ during $\Delta$ (2.4) are constant, then the decomposition of the variance $\Theta(t,t_0)$ (4.10) takes the classical form of Markowitz's variance $\Theta(t,t_0)$ (1.2).

We underline that one should consider the variances of any portfolio in the same way as the variances of any tradable market security. The portfolio variance of prices $\Phi(t)$ (3.18) and the variance of returns $\Theta(t,t_0)$ (3.19) have the same expressions as the variances of prices $\phi_j(t)$ (3.9) and returns $\theta_j(t,t_0)$ (3.10) of any market security $j$. The decompositions of the portfolio variances of prices $\Phi(t)$ (4.4; 4.5) and returns $\Theta(t,t_0)$ (4.10) by its securities are the result of the dependence of time series of the values $Q(t_i)$, volumes $W(t_i)$, and prices $s(t_i)$ (2.11-2.15) that determine the trades with the portfolio as a single market security on the time series that describe the normalized values $c_j(t_i)$, volumes $u_j(t_i)$, and prices $p_j(t_i)$ (2.7- 2.10) of trades of the securities of the portfolio.

The expression of the portfolio variance $\Theta(t,t_0)$ (4.10) highlights that the risks of the portfolio have more complex dependence on the risks of its securities than was described by Markowitz.

## 5. Conclusion

The investor who holds his portfolio and doesn't trade his shares can use the time series of market trades with the securities of the portfolio to assess the portfolio return and variance in the same form as he assesses the return and variance of any market security. The transformations of the time series of market trades with securities that compose the portfolio determine the time series of trades with the portfolio as a single market security. That establishes the equality between the description of any portfolio and any single market security. The decomposition of the portfolio variance by its securities results from the dependence of the portfolio trade time series on the time series of trades with the securities. The decomposition of the variance $\Theta(t,t_0)$ (4.7) is a quadratic form in variables of relative amounts $X_j(t_0)$ invested into securities with the coefficients $\Omega_{jk}(t,t_0)$ (4.8), which, in their turn, are quadratic forms in variables $x_j(t_0)$ (4.2; 4.4). The distinctions from Markowitz's expression of the portfolio variance $\Theta(t,t_0)$ (1.2) are the results of the impact of the random volumes of trades with the securities. If one assumes that all volumes $U_j(t_i)$ of the consecutive trades with all securities of the portfolio during the averaging interval are constant, the variance $\Theta(t,t_0)$ (4.7; 4.8) takes the form (1.2) that was derived by Markowitz. The current methods for selecting the portfolio with higher returns under lower variance that are based on decomposition $\Theta(t,t_0)$ (1.2) are valid only for the approximation that neglects the impact of random trade volumes.

The market-based portfolio selection that accounts for the randomness of the volumes of consecutive market trades is more difficult. The expression of market-based portfolio variance $\Theta(t,t_0)$ (4.7; 4.8) reveals that the dependence of the portfolio risk on the risks of the securities of the portfolio is a more complex problem than it was described by Markowitz (1.2).

To forecast the market-based variance $\Theta(t,t_0)$ (3.19; 4.7; 4.8) of the portfolio at horizon $T$, one should predict the time series of the values and volumes of market trades with all securities of the portfolio at the same horizon $T$ during the averaging interval. Such forecasts seem to be extremely difficult. We consider these tough challenges of predictions of the market trades with securities as the main obstacle for the optimal portfolio selection. In this paper we don't consider these problems.

## Appendix A. Market-based covariances of two securities

The brief derivations of the market-based means and variances of prices and returns use the results (Olkhov, 2022-2025). Let us consider the product (A.1) of two equations (2.5) that describe the trades with values $C_j(t_i)$, $C_k(t_i)$, volumes $U_j(t_i)$, $U_k(t_i)$, and prices $p_j(t_i)$, $p_k(t_i)$ of securities $j$ and $k$ at time $t_i$ during $\Delta$ (2.4):

$$C_j(t_i)C_k(t_i) = p_j(t_i)p_k(t_i)\, U_j(t_i)U_k(t_i) \tag{A.1}$$

We define market-based covariance $\phi_{jk}(t)$ (A.2) of prices $p_j(t_i)$ and $p_k(t_i)$ in the form that is alike to VWAP (3.2; 3.3):

$$\phi_{jk}(t) = \frac{1}{U_{jk}(t)}\frac{1}{N}\sum_{i=1}^{N}\left(p_j(t_i) - p_j(t)\right)\left(p_k(t_i) - p_k(t)\right) U_j(t_i)U_k(t_i) \tag{A.2}$$

$$U_{jk}(t) = \frac{1}{N}\sum_{i=1}^{N} U_j(t_i)U_k(t_i) \tag{A.3}$$

The mean prices $p_j(t)$ and $p_k(t)$ of securities $j$ and $k$ are determined as VWAP (3.2; 3.3). The use of (2.5) transforms (A.2) into:

$$\phi_{jk} = \frac{E[C_j(t_i)C_k(t_i)] - p_k(t)E[C_j(t_i)U_k(t_i)] - p_j(t)E[U_j(t_i)C_k(t_i)]}{E[U_j(t_i)U_k(t_i)]} + p_j(t)p_k(t) \tag{A.4}$$

*E[..]* in (A.4) denotes the assessments of the joint mathematical expectation of the product of values and volumes by $N$ terms of time series, for example:

$$E[C_j(t_i)U_k(t_i)] = \frac{1}{N}\sum_{i=1}^{N} C_j(t_i)U_k(t_i) = C_j(t;1)U_k(t;1) + cov\{C_j(t), U_k(t)\} \tag{A.5}$$

In (A.5) $C_j(t;1)$ and $U_k(t;1)$ denote their averages as in (3.4) and $cov\{C_j(t),U_k(t)\}$ – covariances:

$$cov\{C_j(t), U_k(t)\} = \frac{1}{N}\frac{1}{N}\sum_{i=1}^{N}\left(C_j(t_i) - C_j(t;1)\right)\left(U_k(t_i) - U_k(t;1)\right) \tag{A.6}$$

Let us define functions $\psi_{jk}(t)$, $\chi_{jk}(t)$, and $\varphi_{jk}(t)$ (A.7; A.8):

$$\psi_{jk}(t) = \frac{cov\{C_j(t), C_k(t)\}}{C_j(t;1)C_k(t;1)} \quad ; \quad \varphi_{jk}(t) = \frac{cov\{C_j(t), U_k(t)\}}{C_j(t;1)U_k(t;1)} \tag{A.7}$$

$$\chi_{jk}(t) = \frac{cov\{U_j(t), U_k(t)\}}{U_j(t;1)U_k(t;1)} \quad ; \quad U_{jk}(t) = U_j(t;1)U_k(t;1)[1 + \chi_{jk}(t)] \tag{A.8}$$

If one takes $j=k$, then $\psi_j^2(t)=\psi_{jj}(t)$ and $\chi_j^2(t)= \chi_{jj}(t)$ (A.8) have the meaning of squares of coefficients of variation of the values $C_j(t_i)$ and volumes $U_j(t_i)$ of trades of security $j$ during $\Delta$ (2.4). The functions $\psi_{jk}(t)$, $\chi_{jk}(t)$, and $\varphi_{jk}(t)$ (A.7; A.8) present the covariance $\phi_{jk}(t)$ (A.4) of prices $p_j(t_i)$ and $p_k(t_i)$ as :

$$\phi_{jk}(t) = \frac{\psi_{jk}(t) - 2\varphi_{jk}(t) + \chi_{jk}(t)}{1 + \chi_{jk}(t)}\, p_j(t)p_k(t) \tag{A.9}$$

The covariance $\phi_{jk}(t)$ (A.9) determines the covariance $\theta_{jk}(t,t_0)$ of returns (3.5) of securities $j$ and $k$ with respect to their prices $p_j(t_0)$ and $p_k(t_0)$ at time $t_0$ in the past:

$$\theta_{jk}(t, t_0) = \frac{\phi_{jk}(t)}{p_j(t_0)p_k(t_0)} = \frac{\psi_{jk}(t) - 2\varphi_{jk}(t) + \chi_{jk}(t)}{1 + \chi_{jk}(t)}\, R_j(t, t_0)R_k(t, t_0) \tag{A.10}$$

If $j=k$, the relations (A.9; A.10) describe the variances $\phi_j(t)$ (A.11) of prices $p_j(t_i)$ and the variance $\theta_j(t,t_0)$ (A.11) of returns $R_j(t_i,t_0)$ of security $j=1,..J$ of the portfolio during $\Delta$ (2.4).

$$\phi_j(t) = \frac{\psi_j(t)-2\varphi_j(t)+\chi_j(t)}{1+\chi_j(t)}\ p_j^2(t) \quad ; \quad \theta_j(t,t_0) = \frac{\psi_j(t)-2\varphi_j(t)+\chi_j(t)}{1+\chi_j(t)}\ R_j^2(t,t_0) \qquad \text{(A.11)}$$

## Appendix B. Decomposition of the portfolio variance by its securities

The variance $\Phi(t)$ (3.18; B.1) of prices of the portfolio has the same form as $\phi_j(t)$ (A.11):

$$\Phi(t) = \frac{\psi^2(t)-2\ \varphi(t)+\chi^2(t)}{1+\chi^2(t)}\ s^2(t) \qquad \text{(B.1)}$$

From (A.7; A.8), obtain coefficients of variation $\psi(t)$ (B.2) of the values $Q(t_i)$ and coefficients of variation $\chi(t)$ (B.2) of the volumes $W(t_i)$ of trades with the portfolio as a single security

$$\psi^2(t) = \frac{cov\{Q(t),Q(t)\}}{Q^2(t;1)} = \frac{\Psi_Q(t)}{Q^2(t;1)} \quad ; \quad \chi^2(t) = \frac{cov\{W(t),W(t)\}}{W^2(t;1)} = \frac{\Psi_W(t)}{W^2(t;1)} \qquad \text{(B.2)}$$

$$\Psi_Q(t) = \frac{1}{N}\sum_{i=1}^{N}(Q(t_i)-Q(t;1))^2 = Q(t;2)-Q^2(t;1)\ ;\ \Psi_W(t) = W(t;2)-W^2(t;1) \qquad \text{(B.3)}$$

The function $\varphi(t)$ (B.4) denotes the ratio of the covariance of values and volumes of the portfolio to their mean values $Q(t;1)$ and $W(t;1)$:

$$\varphi(t) = \frac{cov\{Q(t),W(t)\}}{Q(t;1)W(t;1)} \quad ; \quad W(t;2) = W^2(t;1)[1+\chi^2(t)] \qquad \text{(B.4)}$$

$W(t;2)$ (B.4) denoted the mean square of trade volumes $W^2(t_i)$ of the portfolio. Let us use the dependence (2.11) of the values $Q(t_i)$ and volumes $W(t_i)$ of trades with the portfolio on the normalized values $c_j(t_i)$ and volumes $u_j(t_i)$ of trades with securities $j=1,..J$ of the portfolio and substitute (2.11) into (B.2-B.4). Then, obtain:

$$cov\{Q(t),Q(t)\} = \sum_{j,k=1}^{J} cov\{c_j(t_i),c_k(t_i)\}\ ;\ cov\{W(t),W(t)\} = \sum_{j,k=1}^{J} cov\{u_j(t_i),u_k(t_i)\} \qquad \text{(B.5)}$$

$$cov\{Q(t),W(t)\} = \sum_{j,k=1}^{J} cov\{c_j(t_i),u_k(t_i)\} \qquad \text{(B.6)}$$

The relations (B.5; B.6) allow present the decomposition (B.7) of the covariance $\Phi(t)$ (B.1):

$$\Phi(t) = \frac{1}{W(t;2)}\sum_{j,k=1}^{J}[cov\{c_j(t),c_k(t)\} - 2s(t)cov\{c_j(t),u_k(t)\} + s^2(t)cov\{u_j(t),u_k(t)\}] \qquad \text{(B.7)}$$

We use (2.8) and (A.7; A.8) to transform (B.7) and present the decomposition of the covariance $\Phi(t)$ (B.8) of the portfolio by its securities:

$$\Phi(t) = \sum_{j,k=1}^{J}\frac{p_j(t)p_k(t)\psi_{jk}(t) - 2s(t)p_j(t)\varphi_{jk}(t)+s^2(t)\chi_{jk}(t)}{1+\chi^2(t)}\ x_j(t_0)x_k(t_0) \qquad \text{(B.8)}$$

The coefficients $x_j(t_0)$ (4.2) in (B.8) describe the relative numbers of shares $U_j(t_0)$ of the security $j$ in the total number $W_\Sigma(t_0)$ of shares of the portfolio. The decomposition (B.8) hides the dependence of the mean price $s(t)$ (4.2) of the portfolio on the mean prices $p_j(t)$ of securities. Let us substitute $s(t)$ (4.2) into (B.8) and obtain the variance $\Phi(t)$ of prices of the portfolio:

$$\Phi(t) = \frac{1}{1+\chi^2(t)}\left[\sum_{j,k=1}^{J}\psi_{jk}(t)\,p_j(t)p_k(t)\ x_j(t_0)x_k(t_0) - \right.$$

$$-2\sum_{j,k,l=1}^{J}\varphi_{jk}(t)\,p_j(t)p_l(t)\,x_j(t_0)x_k(t_0)x_l(t_0)$$

$$+\sum_{j,k,l,m=1}^{J}\chi_{jk}(t)\,p_l(t)p_m(t)\,x_j(t_0)x_k(t_0)x_l(t_0)x_m(t_0)\,] \qquad \text{(B.9)}$$

The dependence of volumes $W(t_i)$ (2.11) of trades with the portfolio on the normalized volumes $u_j(t_i)$ and (B.5) define the dependence (B.10) of $\chi^2(t)$ (B.2) on the variables $x_j(t_0)$ (4.2) of relative number of shares of security $j$

$$\chi^2(t)=\sum_{j,k=1}^{J}\chi_{jk}(t)\;x_j(t_0)\cdot x_k(t_0) \qquad \text{(B.10)}$$

One can change the indexes in the sums of (B.9) and derive the final form (B.11; B.12) of the variance $\Phi(t)$ of prices of the portfolio:

$$\Phi(t)=\sum_{j,k=1}^{J}\Omega_{jk}(t)\,p_j(t)p_k(t)\,x_j(t_0)x_k(t_0) \qquad \text{(B.11)}$$

The coefficients $\Omega_{jk}(t)$ (B.12) are quadratic forms in the variables $x_j(t_0)$ of relative number of shares of security $j$ in the total number of shares of the portfolio:

$$\Omega_{jk}(t)=\frac{1}{1+\chi^2(t)}\left[\psi_{jk}(t)-2\sum_{l=1}^{J}\varphi_{jl}(t)x_l(t_0)+\sum_{l,m=1}^{J}\chi_{lm}(t)\,x_l(t_0)x_m(t_0)\right] \qquad \text{(B.12)}$$

The use of (B.10) transforms the coefficients $\Omega_{jk}(t)$ (B.12) into (B.13):

$$\Omega_{jk}(t)=1+\frac{1}{1+\chi^2(t)}\left[\psi_{jk}(t)-2\sum_{l=1}^{J}\varphi_{jl}(t)x_l(t_0)-1\right] \qquad \text{(B.13)}$$

The definition of $\varphi_{jk}(t)$ (A.7) and (2.8), give:

$$u_l(t;1)=\frac{U_l(t_0)}{N}\quad ; \quad \frac{U_l(t_0)}{u_l(t;1)\,W_\Sigma(t_0)}=\frac{N}{W_\Sigma(t_0)}=\frac{1}{W(t;1)}$$

$$\sum_{l=1}^{J}\varphi_{jl}(t)x_l(t_0)=\sum_{l=1}^{J}\frac{cov\{c_j(t),u_l(t)\}}{c_j(t;1)u_l(t;1)}\,\frac{U_l(t_0)}{W_\Sigma(t_0)}=\sum_{l=1}^{J}\frac{cov\{c_j(t),u_l(t)\}}{c_j(t;1)W(t;1)} \qquad \text{(B.14)}$$

Let us calculate the sum by index $l=1,..J$:

$$\sum_{l=1}^{J}cov\{c_j(t),u_l(t)\}=\sum_{l=1}^{J}\frac{1}{N}\sum_{i=1}^{N}[c_j(t_i)-c_j(t;1)][u_l(t_i)-u_l(t;1)]=$$

$$=\frac{1}{N}\sum_{i=1}^{N}[c_j(t_i)-c_j(t;1)]\;\sum_{l=1}^{J}[u_l(t_i)-u_l(t;1)] \qquad \text{(B.15)}$$

From (2.10; 2.11; 2.13), obtain:

$$\sum_{l=1}^{J}[u_l(t_i)-u_l(t;1)]=W(t_i)-W(t;1) \qquad \text{(B.16)}$$

From (2.8; B.14 - B.16), obtain:

$$\sum_{l=1}^{J}\varphi_{jl}(t)x_l(t_0)=\omega_j(x)=\frac{cov\{c_j(t),W(t)\}}{c_j(t;1)W(t;1)}=\frac{cov\{C_j(t),W(t)\}}{C_j(t;1)W(t;1)} \qquad \text{(B.17)}$$

$$cov\{C_j(t),W(t)\}=\frac{1}{N}\sum_{i=1}^{N}[C_j(t_i)-C_j(t;1)][W(t_i)-W(t;1)] \qquad \text{(B.18)}$$

Relations (B.17; B.18) present the coefficients $\Omega_{jk}(t)$ (B.13) as (B.19):

$$\Omega_{jk}(t)=\frac{\psi_{jk}(t)-2\omega_j(t)+\chi^2(t)}{1+\chi^2(t)}=1+\frac{1}{1+\chi^2(t)}\left[\psi_{jk}(t)-2\omega_j(t)-1\right] \qquad \text{(B.19)}$$

The relations (B.11; B.19) and (A.7; A.8; B.10; B.17) determine the decomposition of the variance $\Phi(t)$ of prices by the securities $j=1,..J$ of the portfolio.

To derive the decomposition of the market-based variance $\Theta(t,t_0)$ (3.19) of returns of the portfolio by its securities we use (B.11; B1.9):

$$\Theta(t,t_0)=\frac{\Phi(t)}{s^2(t_0)}=\sum_{j,k=1}^{J}\Omega_{jk}(t)\frac{p_j(t)}{p_j(t_0)}\frac{p_j(t_0)x_j(t_0)}{s(t_0)}\frac{p_k(t)}{p_k(t_0)}\frac{p_k(t_0)x_k(t_0)}{s(t_0)} \tag{B.20}$$

The definition of average return $R_j(t,t_0)$ (3.7) of security $j$ and the definitions $X_j(t_0)$ (4.3) of the relative amount invested into security $j$ of the portfolio at time $t_0$, transform the variance $\Theta(t,t_0)$ (B.20) of returns of the portfolio into:

$$\Theta(t,t_0)=\sum_{j,k=1}^{J}\Omega_{jk}(t)\,R_j(t,t_0)R_k(t,t_0)X_j(t_0)X_k(t_0) \tag{B.21}$$

The expression (B.21) describes the decomposition the variance $\Theta(t,t_0)$ of returns of the portfolio by its securities as a quadratic form in the variables $X_j(t_0)$ of relative amounts invested into security of the portfolio at time $t_0$. The coefficients $\Omega_{jk}(t)R_j(t,t_0)R_k(t,t_0)$ (B.19; B.21) of the quadratic form (B.21) differ a lot from the covariance $\theta_{jk}(t,t_0)$ (1.3) of securities $j$ and $k$.

If one assumes that the volumes of trades $U_j(t_i)$ with all securities $j=1,..J$ of the portfolio during $\Delta$ (2.4) are constant $U_j(t_i)=U_j$, then the volumes $W(t_i)=W$ of trade with the portfolio are constant. From (3.8), obtain that $\chi_{jk}(t)=0$ and from (B.17), obtain that $\omega_j(t)=0$. For the assumption that all trade volumes $U_j(t_i)=U_j$ are constant, from (B.19; B.21) and (A.7), obtain:

$$\Omega_{jk}(t)R_j(t,t_0)R_k(t,t_0)=\psi_{jk}(t)R_j(t,t_0)R_k(t,t_0)=\frac{cov\{C_j(t),C_k(t)\}}{C_j(t;1)C_k(t;1)}\frac{p_j(t)}{p_j(t_0)}\frac{p_k(t)}{p_k(t_0)}$$

We use relations (2.15) and for constant trade volumes $U_j$, obtain:

$$\frac{cov\{C_j(t),C_k(t)\}}{C_j(t;1)C_k(t;1)}=\frac{U_jU_k cov\{p_j(t),p_k(t)\}}{U_jU_k p_j(t)p_k(t)}$$

From above, obtain:

$$\Omega_{jk}(t)R_j(t,t_0)R_k(t,t_0)=\frac{cov\{p_j(t),p_k(t)\}}{p_j(t_0)p_j(t_0)}=cov\{R_j(t,t_0),R_k(t,t_0)\}=\theta_{jk}(t,t_0) \tag{B.22}$$

In (B.22) the covariance $\theta_{jk}(t,t_0)$ coincides with the covariance (1.3) of securities $j$ and $k$. Finaly, obtain that if one assumes that all trade volumes $U_j(t_i)=U_j$ with all securities $j=1,..J$ of the portfolio are constant, the variance $\Theta(t,t_0)$ (B.21) of returns of the portfolio coincides with Markowitz's expressions (1.2; 1.3) of the portfolio variance:

$$\Theta(t,t_0)=\sum_{j,k=1}^{J}\theta_{jk}(t)\,X_j(t_0)X_k(t_0)$$